\documentclass[aps,pre,twocolumn,superscriptaddress,amsmath,amssymb,floatfix,showpacs]{revtex4}
\usepackage{graphicx}
\usepackage{times}
\usepackage{natbib}
\bibpunct{[}{]}{,}{n}{,}{,}
\begin{document}
\def\d{\delta}
\def\D{\Delta}
\def\s{\sigma}
\def\g{\gamma}
\def\e{\epsilon}
\def\b{\beta}
\def\a{\alpha}
\def\l{\lambda}
\def\L{\Lambda}

\title{Propagation of fluctuations in interaction
networks governed by the law of mass action.}
\author{Sergei Maslov
}
\affiliation{Department of Condensed Matter Physics,
Brookhaven National Laboratory, Upton, New York 11973, USA}
\altaffiliation{Corresponding author, FAX (631) 344-2918}\email{maslov@bnl.gov}
\author{Kim Sneppen}
\affiliation{Niels Bohr Institute,
University of Copenhagen,
Blegdamsvej 17, 2100 Copenhagen \O, Denmark}
\author{I. Ispolatov}
\affiliation{Ariadne Genomics Inc., 9700 Great Seneca Highway, Suite 113,
Rockville, Maryland 20850, USA}
\altaffiliation{Corresponding author, FAX (240) 453-9026. Permanent address:
Departamento de Fisica, Universidad de Santiago de Chile,
Casilla 302, Correo 2, Santiago, Chile}
\email{iispolat@lauca.usach.cl}
\begin{abstract}
Using an example of physical interactions between proteins,
we study how perturbations propagate in interconnected networks
whose equilibrium state is governed by the law of mass action.
We introduce a comprehensive matrix formalism which predicts
the response of this equilibrium to
small changes in total concentrations of individual molecules,
and explain it using a heuristic
analogy to a current flow in a
network of resistors. Our main conclusion is that on average
changes in free concentrations exponentially decay
with the distance from the source of perturbation. We then study
how this decay is influenced by such factors as
the topology of a network, binding strength,
and correlations between concentrations of neighboring nodes.
An exact analytic expression for
the decay constant is obtained for the case of
uniform interactions on the Bethe lattice.
Our general findings are illustrated using a real biological
network of protein-protein interactions in baker's yeast
with experimentally determined protein concentrations.
\end{abstract}

\maketitle

Equilibria in a broad class of microscopically reversible processes
whose direct and reverse rates are proportional to the product of
concentrations of participating substances is described by
the Law of Mass Action (LMA). It has been rigorously proven that the unique
steady state of such processes is completely defined by the
set of initial concentrations and reaction constants \cite{Shear68}.
Recently, it has become popular to
visualize large sets of interacting substances as networks, where nodes and
links correspond to the reactants and their propensity for pairwise interactions.
One of the best-studied examples of such networks is that formed
by all protein-protein physical interactions (pairwise bindings)
in a given organism. The LMA determines the equilibrium free concentrations
of individual proteins as well as those of hetero- and homo- dimers.
A surprising feature observed in virtually all recent large-scale studies of
these networks in a wide-ranging variety of biological organisms is their
globally connected topology. Indeed, most pairs of protein nodes
are linked to each other by relatively short chains of interactions.
Total (bound plus unbound) concentrations
of individual proteins are subject to both stochastic fluctuations
due to the noise in their production and degradation as well as
to systematic changes in response to external and internal stimuli.
Such localized perturbation changes bound and free concentrations of
immediate neighbors of the perturbed protein which in their turn influence
their neighbors, etc.
Thus a fluctuation in the total concentration of just one reactant
to some degree affects free concentrations of
all substances in the same connected component of the
network. The propagation of fluctuations far away from their source
presents a great threat of undesirable cross-talk between different functional
processes, simultaneously taking place in an organism. Thus it is important to understand
whether and how this propagation gets attenuated
and under what conditions it is
minimized. There also exist several anecdotal cases when
changes in free concentrations propagating beyond the immediate neighbors of
a perturbed protein are used for a meaningful biological regulation/signaling.
Then a relevant question is under what conditions the
propagation of the signal in a desirable direction is the least attenuated.

In this work we study how localized perturbations such as changes of  concentrations
of individual proteins affect the binding equilibrium at all nodes of
a protein interaction network. Our
main conclusion is that under a broad range of conditions such perturbations
exponentially decay with the network distance away from the perturbed node.
Luckily, this makes protein binding networks
poor conduits for indiscriminately propagating fluctuations
which would have led to undesirable cross-talk between biologically distinct pathways.
On the other hand, under carefully selected conditions, a perturbation can
propagate relatively far with a minimal attenuation.
We first consider a general
case of propagation of small perturbations in a network of an arbitrary topology,
concentrations, and dissociation constants.
Several simpler
analytically tractable case studies follow.

Consider a network of pairwise interactions between  $N$ distinct types of
proteins (or any other molecules for that matter), where each
protein corresponds to a vertex. The existence of a link between vertices
$i$ and $j$ means that substances $i$ and $j$ interact to form a bound
complex (hetero- or homo- dimer)
$ij$. Throughout this paper we would consider only such dimers
and ignore the existence of multi-protein complexes consisting of 3 or more
proteins. However, we checked \cite{upcoming} that our main results could be
easily extended to an arbitrary composition of complexes or even to a
situation in which individual nodes combine with each other
in reversible chemical reactions obeying the LMA.

The LMA dictates that free concentrations of proteins $F_i$ and those of dimers $D_{ij}$
obey
\begin{equation}
\label{lma}
  {F} _i \; {F} _j=k_{ij} {D}_{ij},
\end{equation}
where $k_{ij}$ is the corresponding dissociation constant.
%
Adding the mass conservation one obtains the following system of
equations that relates total concentrations $C_i$ of proteins to their
free concentrations $F_i$,
\begin{equation}
\label{main}
C_i=F_i + \sum_{j \leftrightarrow i} D_{ij}= F_i+\sum_{j \leftrightarrow i}
\frac { F_i F_j } {k_{ij}} \qquad .
\end{equation}
Here and below the notation $\sum_{j \leftrightarrow i}$ means a sum over all
vertices $j$ that are the network neighbors of the vertex $i$.
In a general case of 4 or more interconnected interacting pairs
this system of non-linear equations allows for only numerical
solution. One
particularly convenient computational method involves rewriting the Eq.
(\ref{main}) as
\begin{equation}
\label{iter}
F_i=\frac {C_i}{1 + \sum_{j \leftrightarrow i}  F_j/k_{ij}}
\end{equation}
and successively iterating it starting with $F_i=C_i$
until the LMA is satisfied with a desired precision.
The proof presented in Ref. \cite{Shear68}
guarantees the uniqueness of the solution found this way.
Now consider a change in free concentrations $\d F_j$
induced by a small perturbation of total concentrations $\d C_i$.
Linearization of Eq.~(\ref{main}) yields
\begin{equation}
\label{small}
\frac{\d C_i} {C_i} = \sum_j \L _{ij} \frac { \d F_j} { F_j} \qquad ,
\end{equation}
with matrix $\hat \L$ \cite{homodimer_caveat} given by
\begin{equation}
\label{matrix}
\L_{ij}=
A_{ij}\frac{D_{ij}}{C_i}+\d_{ij} \qquad .
\end{equation}
Here $A_{ij}$ is the adjacency matrix of the network.
In case of multi-protein complexes consisting of more than 2 different
proteins $D_{ij}$ should be replaced
with the total concentration of all complexes containing both $i$
and $j$ among their constituents.
It follows from Eqs.~(\ref{small},\ref{matrix})
that if the change in total concentration was limited to just one node $0$,
the induced relative change of free concentration of any other node $i \neq 0$
satisfies
\begin{equation}
\label{neighbor}
\frac {\d F_i} {F_i}=-\sum_{j \leftrightarrow i} \frac {D_{ij}}
{C_i} \frac {\d F_j} {F_j} \qquad .
\end{equation}
This equation shows that changes in free concentrations on nearest
neighbors tend to be of the opposite sign. Also, since
$\sum_{j \leftrightarrow i} D_{ij}/C_i=1-F_i/C_i<1$, the
absolute magnitude of perturbation $|\d F_i/F_i|$ on any node away
from the source is less or equal than its maximal value among its neighbors:
$\max_{j \leftrightarrow i} |\d F_j/F_j|$.
Bonds with higher $D_{ij}/C_i$ are better transmitters
of perturbations from node $j$ to node $i$.
Note, that this quantity is non-symmetric: The
transmission along any particular edge is directional
with preferred direction pointing from the higher total concentration to lower
one.

Inverting (\ref{small}) one obtains the desired expression for the linear response
of any free concentration to an arbitrary perturbation in total
concentrations:
\begin{equation}
\label{matrix_inv}
\frac{\d F_i} {F_i} = \sum_j \left({\hat \L}^{-1}\right)_{ij} \frac{\d
C_j}{C_j}\qquad .
\end{equation}
%
\begin{figure}
\includegraphics[width=.45\textwidth]{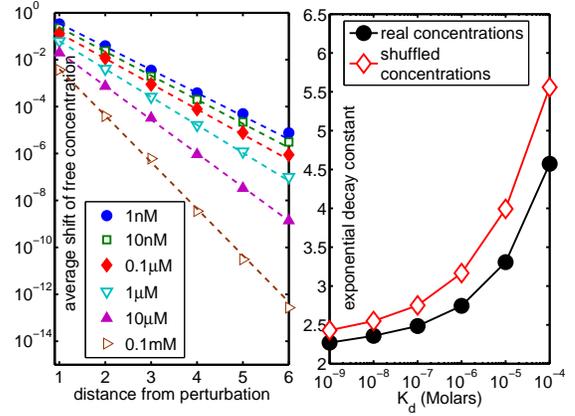}
\caption{\label{fig1}
A) The average magnitude of normalized changes in free concentrations
$|\d F_i/F_i|$ per unit of $\d C_0/C_0$ plotted as a function of the distance
$l_{i0}$ from the perturbed node $0$. The propagation of perturbations
(Eq. \ref{matrix_inv}) was
computed in a highly curated yeast PPI network \cite{BIOGRID} with
real-life total concentrations of individual proteins
\cite{yeast_conc} and dissociation constant
$K_d$ increasing in 10-fold increments starting from
$1$nM (or 34 molecules/cell)
and up to 0.1mM (the steepest decaying curve).\\
B) The exponential decay constant $\gamma (K_d)$ obtained
from the best fit in the form
$A\exp[-\gamma (K_d) l_{i0}]$ to curves shown in the Panel A with
the real-life concentrations
(filled circles), and for randomly reshuffled
concentrations (open diamonds).
}
\end{figure}
As an application of the general theory outlined above
we calculated the free concentrations $F_i$
and the linear response matrix ${\hat \L}^{-1}$
in a realistic protein network of baker's yeast
{\it Saccharomyces cerevisiae}. A highly-curated
set of protein-protein physical interactions from the
BIOGRID dataset \cite{BIOGRID} was further restricted to include
only interactions which were reported in at least two publications.
Total concentrations of proteins for yeast grown in rich growth medium
conditions were taken from a genome-wide experimental
study \cite{yeast_conc}.
The resulting dataset consists of  4185 interactions among
1740 proteins with total concentrations ranging between 50 and
1 million molecules/cell with median $\sim$ 3000
molecules/cell.
In the absence of
genome-wide information regarding the value of dissociation
constants in our simulations we assume them all to be the same $k_{ij}=K_d$.
We also tried simulations in which $k_{ij}$ are
randomly drawn from a particular probability
distribution. In all of our simulations we observed that
the magnitude of relative changes in free concentrations
exhibits a universally exponential decay with the network distance
from the source of perturbation
(Fig.~\ref{fig1}A). That is to say, on average, matrix elements of
$\left(\L^{-1}\right)_{ij}$ exponentially decrease as a function of
distance between $i$ and $j$. Further progress in experimental methods will
evidently lead to discovery of new protein interactions and more precise
values of protein concentrations and dissociation constants. However, such
improvements will not affect our general conclusion about the
exponential decay of perturbations.

While the linear response approximation (\ref{matrix_inv})
describes infinitesimally small
perturbations, the response to
a finite perturbation could be obtained numerically
by repeating iterations
of (\ref{iter}) with perturbed total concentrations
and comparing the resulting free concentrations to the
wild-type ones. We simulated the response of the system to
a 20\% decrease (which is roughly the range of intrinsic
fluctuations \cite{newman2006scp}) as well as to a complete elimination
(which is experimentally realizable as a gene knock-out or inactivation)
for each of the proteins in our dataset. It was found that
even in the latter extreme case the linear response approximation works
rather well. The exponential decay of perturbation was found to be
identical to that calculated using the Eq. (\ref{iter}) and
the overall magnitude of
changes in free concentration is comparable to that predicted
from the linear response.

To develop a better understanding of this matrix formalism,
we first consider the case when the underlying network is
bipartite (but not necessarily acyclic).
To take into account the natural sign-alternation of $\d F_i/F_i$
on immediate neighbors in the network (see Eq.~(\ref{neighbor})), we
introduce the new variables
$\phi_i=(-1)^{x_i} \d F_i/F_i$ where $x_i$ is 0
on one sublattice and 1 on the other.
This allows us to rewrite the
Eqs.~(\ref{small},\ref{matrix}) as
\begin{equation}
\label{kirchoff_law}
\d \tilde{C}_i  = \sum_j \sigma_{ij} \phi_j \qquad .
\end{equation}
Here $\d \tilde{C}_i=(-1)^{x_i} \d C_i$ and $\hat \sigma$ is given by
\begin{equation}
\label{rhoij}
\sigma_{ij}=
-A_{ij}D_{ij}+\delta_{ij} \left(\sum_{k \leftrightarrow i} D_{ik}+F_i \right ) \qquad .
\end{equation}
%
In the situation when
$\d \tilde{C}_i=\delta_{i0} \d C_0$ (i.e. when the
perturbation is limited to a single node $0$),
%
the Eqs.(\ref{kirchoff_law},\ref{rhoij})
can be interpreted
as describing ``electric potentials'' $\phi_i$ in
the network of resistors with
resistances $R_{ij}=1/\sigma_{ij}=1/D_{ij}$ subject to
the injection of the current $\d C_0 $ at the node $0$.
Each node is also shunted to an auxiliary ``ground node''
with potential $\phi_{G}=0$ by the resistance $R_{iG}=1/F_i$.
Potential gradients along edges
$\phi_i-\phi_j=(-1)^{x_i} \d F_i/F_i-(-1)^{x_j} \d F_j/F_j=
(-1)^{x_i} \d D_{ij}/D_{ij}$
determine relative (dimensionless) changes in concentrations of heterodimers,
while currents $I_{ij}=(\phi_i-\phi_j)/R_{ij}=(-1)^{x_i} \d D_{ij}$ --
the absolute (dimensional) changes.
Similarly, currents to the ground
$I_{iG}=\phi_i/R_{iG}=(-1)^{x_i} \d F_i$ are equal to changes in
free concentrations of proteins.
As in resistor networks, the Kirchoff Law here follows from the
mass conservation which states that everywhere the total current
flowing out of node $i$ -
$I_{iG}+\sum_{j \leftrightarrow i} I_{ij}=
(-1)^{x_i}(\d F_i +\sum_{i \leftrightarrow j}\d
D_{ij})=(-1)^{x_i}\d C_i=\d \tilde{C}_i$
is equal to the external current $\d \tilde{C}_i=\d _{i0} \d C_0$
of changes in total concentrations.

The interpretation of the free concentrations $F_i$ as ``shunt conductivities''
leaking the ``current'' to the ground means that the smaller they are,
the weaker is the decay of
both currents $\d D$ and $\d F$ with the distance.
Since stronger binding generally decreases
free concentrations of all proteins, it naturally
reduces the rate of decay of perturbations (visible in the
Fig. \ref{fig1}A). However, the exponential decay constant
$\gamma (K_d)$ appears to saturate around $2.25$ as
$K_d \to 0$ (Fig. \ref{fig1}B).
%
%
%

This saturation is easy to understand.
Indeed, consider the most ideal scenario in which all
free concentrations $F_i$ are very small and thus the ``current'' $\d D$
is approximately conserved (loss to the ground is negligible.)
The exponential growth in the number of neighbors
$N_{n}(l) \sim (\langle d(d-1) \rangle/\langle d \rangle )^l$
as a function of distance $l$ from the perturbation source means that
even in this ideal setup the average current at distance $l$ would be proportional to
$1/N_{n}(l)$ and thus exponentially small. The same rate of decay
describes the ``potentials'' $\d F_i/F_i$ as well.

\begin{figure}
\includegraphics[width=.35\textwidth]{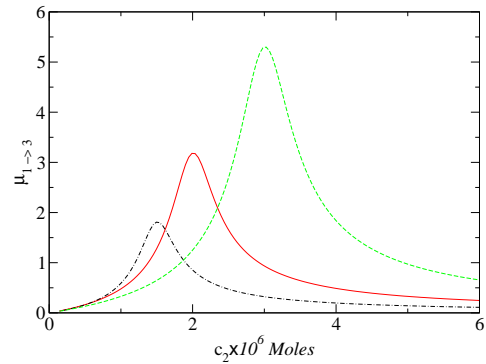}
\caption{\label{fig_3}
The propagation ratio $\mu_{1 \to 3}$ defined in the text
in a linear chain $C_1 \to C_2 \to C_3$ with strong binding ($k=1$nM)
as a function of concentration of
the intermediate substance $C_2$. All curves have the same
concentration of the target protein $C_3=1\mu$M and different
concentrations of the source of the signal $C_1=2\mu$M (dashed
line), $C_1=1\mu$M (solid line), and $C_1=0.5\mu$M (dot-dashed
line).
In the strong binding
limit the propagation ratio reaches its maximum when all three
substances are completely bound, $C_2=C_1+C_3$. It also follows that the
peak propagation is better from high to low total concentration i.e. when
$C_1>C_3$.}
\end{figure}

However, it is important to emphasize that the ideal scenario
outlined above is almost never realized with real-life
concentrations. Indeed, the limit of infinitely strong
binding $k_{ij} \to 0$ does not make {\it all} but only {\it some} of free
concentrations $F_i$ to go to zero. One can see it clearly
already for two interacting proteins. When their concentrations
$C_1$ and $C_2$ are not equal to each other,
in the strong binding limit $k_{12} \to 0$ the free concentration
of the more abundant protein (say 1) remains nonzero
$F_1 \to C_1-C_2$, while the free concentration of its less abundant
partner $F_2 \to 0$. Consider another simple
example of a chain of three proteins with initial
concentrations $C_1$, $C_2$ and $C_3$ reacting to form
dimers $1-2$ and $2-3$
with the same dissociation constant $k$. In this case
one could still analytically calculate all free and bound
concentrations.
The logarithmic derivative $\mu_{1 \to 3}=(\partial F_3/\partial C_1) (C_1/F_3)$
quantifies the propagation of perturbation of the node $1$ through this 3-node
channel and in the strong binding limit has a maximum around
$C_2=C_1+C_3$ (see Fig.\ref{fig_3}) i.e. when
all three substances are completely bound: $F_1,F_2,F_3 \to 0$.
In a general case of a PPI network of arbitrary
topology the only situation in which free concentrations of all proteins
would approach zero as $k_{ij} \to 0$ is when their
total concentrations $C_i$ are proportional to their degrees
$d_i$. For a given topology of the network
such concentration setup has the slowest decay of
perturbations.
%

Most real-life PPI networks are characterized by a positive
correlation between total concentrations of interacting proteins.
In the yeast network used in this study we observed this effect to be present and
highly statistically significant: (the Spearman rank correlation
coefficient of 0.27 with P-value of
$10^{-54}$.) Such correlation improves the balance
between total concentrations of interacting nodes and thus
somewhat lowers the average free concentration of proteins
compared to a case where this correlation is absent. Based
on this we expect that real protein-protein networks
would be more prone to propagating perturbations than their counterparts
in which concentrations of proteins are randomly
reshuffled and thus the correlation between concentrations of
interacting nodes is destroyed. This theoretical expectation was indeed
verified in Fig. \ref{fig1}B (compare filled circles and open diamonds
for the network with real concentrations and the reshuffled ones).
%

Most real-life PPI networks are not bipartite. Fortunately,
due to a relative sparsity of links the number of
odd-length loops in them is small and our
resistor network analogy provides a reasonable
approximation.
For any starting point of perturbation $0$ the optimal
way to define sign-alteration in variables $\phi_i=(-1)^{x_i} \d
F_i/F_i$ is by using $x_i=l_{i0}$ (here $l_{i0}$ is the
distance from the source of perturbation $0$ to
the node $i$.) The majority of links would connect nodes
with opposite 'parities', while
the remaining non-bipartite links could be treated as
a small but important correction to the ideal case.
Like shunt conductivities to the ground,
they contribute to the dissipation of the ``current''.
Indeed, if a link $i \leftrightarrow j$ is of this anomalous kind,
its contribution to the current leaving nodes $i$ and $j$
are equal to each other and given by
$D_{ij}(\phi_i+\phi_j)$. One example of such anomalous
(non-bipartite) links is given by homodimers (see
\cite{homodimer_caveat}.)
In general, these anomalous links lead to the loss
of the total current from the system and thus
tend to dampen the propagation of perturbations.

In what follows we analytically investigate a
simple example of a bipartite network
 -- the Bethe lattice -- where all dissociation constants
 are the same, $k_{ij}=k$, and
each vertex
has the same number of interaction partners (degree) $d_i=d$. When
total concentrations of all proteins are also identical $C_i=C$, their
equilibrium free concentrations are given by
\begin{equation}
\label{mds}
F_i =F=\frac{k}{2d} \left( \sqrt{1 + \frac {4 d C} {k}} - 1 \right).
\end{equation}
%
Small deviations $\d F_l$ at distance $l$ from the perturbation
source at node $0$ should satisfy the Eq. (\ref{neighbor}), which in this
case could be rewritten as
\begin{equation}
\label{rec}
- (d+\frac{k}{F})\d F_l=(d-1)\d F_{l+1}+ \d F_{l-1} \qquad .
\end{equation}
This equation has an exponentially decaying solution $\d F_l=\d F_0 \; \l^l$,
where
\begin{equation}
\label{lambda_d}
 \l =-\frac {1} {d-1} \left(\frac {d+k/F}{2}-\sqrt{\left(\frac
 {d+k/F}{2}\right)^2-(d-1)} \right).
\end{equation}
As expected, $-1<\l<0$ which means that perturbations
sign alternate and exponentially decay
as a function of $l$.
In a strong binding limit the combination of Eqs.
(\ref{lambda_d},\ref{mds}) yields
\begin{equation}
\label{lamnl}
\l=-\frac{1}{d-1}\left(1- \frac{1}{d-2}\sqrt{\frac{dk}{C}} \right)+ {\cal O}
(\frac{dk}{C}),\qquad .
\end{equation}
For a linear chain of proteins ($d=2$), the complete
solution in terms of $C$ and $k$ looks particularly
elegant:
\begin{equation}
\label{lam}
\l_{L}=- \frac{ (1+8C/k)^{1/4}-1}{ (1+8C/k)^{1/4}+1}.
\end{equation}
In the limit of strong binding, a
perturbation in a linear chain propagates indefinitely,
$|\l_{1D}| \rightarrow 1$.
Using this approach,
it also turns possible to find the exact solution for the decay exponent
in the linear chain ($d=2$) with oscillating total concentrations of
proteins:
$C_i= C [ 1 + (-1)^i a]$.
Response of the
even- and odd-numbered vertices has different amplitudes
yet decays with
the same exponential coefficient
%
%
%
$
 \l_{\pm}=-[4\chi\sqrt{1-a^2}-\sqrt{2f(1+4\chi)-2f^2}]/[1+4\chi-f],
$
where $\chi=C/k$ and $f=\sqrt{16 a^2 \chi^2 +8 \chi +1}$.
Evidently,
$|\l_{\pm}(C/k,a)| \leq \l_{L}(C/k)$ with equality being
achieved only for $a=0$. Thus, as was argued in a general
case, any variation in $C_i/d_i$
results in a
larger average unbound concentrations $F_i$ and thus
in a faster decay of perturbations.


This work was supported by 1 R01 GM068954-01 grant from the NIGMS.
Work at Brookhaven National Laboratory was carried out under
Contract No. DE-AC02-98CH10886, Division of Material Science, U.S.
Department of Energy. Work at the NBI was
supported by the Danish National Research Foundation.
II and KS thank Theory Institute for Strongly Correlated and
Complex Systems at BNL for financial support during their
visits.



\begin{thebibliography}{99}
\bibitem{homodimer_caveat} Above we implicitly assumed that every protein
enters every dimer in one copy. That is obviously not true for
homodimers (as well as possibly for some heterodimers). For
proteins forming homodimers the conservation law \ref{main} changes to
$C_i=F_i+2F_i^2/k_{ii}+\sum_{j \neq i} F_iF_j/k_{ij}$ and thus
the diagonal term is given by
$\Lambda_{ii}=1+2D_{ii}/C_i$.
\end{thebibliography}

\end{document}